# Understanding Stigmatizing Language in Clinical Documentation: A Paired Comparison of Ambient AI Drafts and Clinician Finalized Notes

**Yiliang Zhou, MS[1], Yawen Guo, MISM[1], Sairam Sutari MS[2], Jasmine Dhillon BS[5], Alexandra L. Beck BS[5], Emilie Chow MD[3], Steven Tam MD[3], Danielle Perret MD[4], Deepti Pandita MD[3], Gelareh Sadigh MD[5], Archana J. McEligot PhD[6], Kai Zheng, PhD[1]**
**[1]Department of Informatics, University of California, Irvine, Irvine, CA, USA;**
**[2]Institute for Clinical and Translational Science, University of California, Irvine, Irvine, CA, USA;**
**[3]Department of Medicine, University of California, Irvine, Irvine, CA, USA;**
**[4]Department of Physical Medicine & Rehabilitation, University of California, Irvine, Irvine, CA, USA**
**[5]Department of Radiological Sciences, University of California, Irvine, Irvine, CA, USA;**
**[6]Department of Public Health, California State University Fullerton, Fullerton, CA, USA;**

**Abstract**
*Ambient artificial intelligence (AI) documentation tools are increasingly deployed to reduce clinician documentation burden, but their implications for biased language in clinical notes remain unclear. We conducted a large-scale comparison analysis of AI drafts and corresponding clinician finalized notes to quantify stigmatizing language changes pre- and post-editing. Using a lexicon-based natural language processing (NLP) pipeline, we measured (1) the prevalence of stigmatizing language in AI drafts, (2) the prevalence and term composition in final notes, and (3) the frequency of removal or introduction of stigmatizing terms. Across 66,297 paired note sections, 21.4% of AI draft sections contained at least one stigmatizing language mention, rising to 24.0% in clinician finalized versions. Introductions occurred more often than removals, suggesting clinician editing can be a net source of stigmatizing language entering the EHR with using Ambient AI.*

## Introduction

Disparities persist in the United States healthcare system [1], where implicit biases, racism, and discrimination are the major contributors to these inequities [2,3]. Stigmatizing language (SL) may appear through clinical practice and can further perpetuate these disparities. The National Institute on Drug Abuse (NIDA) defines stigma as discrimination directed toward an identifiable group of people, a place, or a nation, and such stigma can operate at interpersonal, institutional, and structural levels [4,5]. Ryan et al. found weight-stigmatizing terms remain prevalent within interpersonal interactions between patients-provider interactions and can subtly undermine patients' ability to obtain timely, equitable access to needed healthcare services [6]. Kelly et al. similarly showed that how individuals with substance-related conditions are labeled can shape perceptions, with the term "abuser" can unintentionally evoke and reinforce stigmatizing attitudes [7]. Several other studies further demonstrated that physicians' electronic health records (EHRs) documentation practice differs by patients' racial and ethnic background, with notes for some minority groups showing more negative affect and fewer positive, trust- and joy-related terms, consistent with prior work linking differential and more stigmatizing documentation to adverse and inequitable health outcomes [2,4,8–11].

Beginning in 2009, the Health Information Technology for Economic and Clinical Health (HITECH) Act aimed to accelerate nationwide adoption of EHRs and other health IT through federal incentives and standards, with the goal of improving healthcare quality, safety, and efficiency, while strengthening privacy and security protections and laying the infrastructure for model digitally enabled healthcare system [12,13]. Additionally, the 21st Century Cures Act established a legal framework to improve how health data are accessed, used, and exchanged through health information technology, and granted patients access to their EHRs via patient portals, allowing them to view clinicians' notes [14–16]. Stigmatizing language is pervasive in clinical notes, and its presence may adversely affect patient care by reinforcing provider bias, shaping clinical decision-making through preconceived assumptions, and potentially eroding trust between patients and clinicians [2–4,8,11,16–24].

Building on the foundation of widespread EHR adoption, ambient artificial intelligence (AI) documentation tools have emerged as using automatic speech recognition (ASR) and generative AI (GenAI) to convert patient–clinician conversations into clinical notes [25–28]. Existing studies suggest that ambient AI documentation tools may improve clinician workflow and reduce documentation burden [26,29–32]. Afshar et al. reported reductions in work exhaustion and interpersonal disengagement, although professional fulfillment did not significantly improve [25]. Guo et al. found that using ambient AI for note drafting was associated with less time spent on documentation and longer notes [27]. Duggan et al. similarly observed operational gains, including a 9.3% higher same-day encounter closure rate and roughly 30% less after-hours work time per day [33].

Despite these reported efficiency gains, current studies also began to evaluate the content of ambient AI–generated documentation. Anderson et al. reported that transcripts produced by ambient digital scribe platforms contained an average of 13.9 errors per note, and that 19.5% of transcript errors propagated into the resulting clinical note [34]. Lukac et al. similarly found that experts rated clinically significant inaccuracies as occurring "occasionally" [29]. Even so, the literature remains silent on whether ambient AI faithfully and neutrally scribes in documentation, and whether biased outputs shape what ultimately enters the EHR. To address this gap, we quantitatively examine SL in ambient AI draft notes and in subsequent clinician finalized documentation, focusing on four key aspects: (1) the extent to which ambient AI versions contain SL, (2) how clinician editing changes the prevalence and composition of SL from AI drafts to final EHR notes, (3) the frequency with which SL present in AI drafts is removed, and (4) the extent to which clinicians introduce new SL when editing AI drafts.

## Materials and Methods

This study was approved by the University of California, Irvine Institutional Review Board (UCI IRB #7123), and a waiver of consent and HIPAA authorization was obtained.

### *Dataset Description*

Between late 2023 and mid 2025, two commercial ambient AI documentation platforms were piloted in outpatient clinics of University of California Irvine Health. Vendors are referred to as Vendor A and Vendor B to preserve contractual and institutional confidentiality. Clinicians can selectively choose up to four AI-assisted note sections to insert into a note, including History of Present Illness (HPI), Physical Exam, Assessment & Plan (A&P), and Results. Clinicians can then edit the generated text before signing off in the EHR. We extracted 98,331 pairs of clinical note sections and its AI draft (Vendor A: 74,133, Vendor B: 23,198) from EHR. For each note section pair, the source data included (1) the ambient AI draft text and (2) the corresponding clinician finalized note text. This section level paired ambient AI draft and clinician finalized dataset was developed in a prior content analysis study [35]. Section-level pairs in which the clinician finalized text was empty (missing or whitespace-only) were excluded.

### *Stigmatizing Lexicon*

To identify SL in note text, we used a domain lexicon of 251 stigmatizing language (SL) terms that was developed in a prior study[36] ,which compiled from existing SL lexicons reported in the literature. The lexicon includes single-word and multiword expressions commonly used to describe patients in ways that may be judgmental, blame-oriented, or otherwise stigmatizing. In this study, we treated the presence of any lexicon term in text as a candidate SL instance and applied consistent preprocessing to support robust matching across documentation styles, including case normalization and handling of morphological variants. Multiword terms were matched as contiguous phrases. We subsequently applied text assertion and contextual classification with natural language processing (NLP) techniques.

### *Clinical NLP Pipeline for SL Assertion*

We implemented a rule based clinical NLP pipeline using SciSpaCy to extract SL mentions from note text and label their assertion status [37]. We used NLTK to compute sentence and word counts for each note section [38]. SL detection was performed with an EntityRuler configured for case insensitive phrase matching, using patterns derived from the lexicon and labeling matched spans as Stigma entities. The pipeline was applied at the section level to both ambient AI drafts and clinician finalized text, and for each section we derived summary measures including sentence count, word count, presence of any SL, total number of SL mentions, and the extracted term level mentions.

***SL Prevalence and Density Analyses***

We summarized the occurrence of SL in both ambient AI drafts and clinician finalized sections using the proportion of sections containing at least one stigmatizing expression. We characterized overall SL density using both raw mention counts and length-standardized rates per 200 words, reflecting the average section length in our dataset. We report point estimates with 95% confidence intervals for proportions and distributional summaries for counts and rates.

To quantify how clinician editing altered SL from draft to final, we assessed changes in presence using paired comparisons of whether any SL appeared in each version. We reported the four paired states (present only in the draft, present only in the final, present in both, present in neither), and used McNemar's test to evaluate whether the probability of containing any SL differed between drafts and final notes under the paired design. We further assessed changes in SL density by comparing paired length-standardized mention rates between the draft and final text using the Wilcoxon signed-rank test. We summarized the distribution of paired density rate changes within sections where SL appeared in either version to reduce dilution from pairs with zero mentions in both versions. A significance level of 0.05 was utilized for the statistical tests, and all analyses were conducted in *R* version 4.1.3.

***Frequencies of SL Addition and Deletion***

For term-level characterization of deletions and additions, we extracted stigmatizing terms in both AI drafted and clinician finalized note sections. For each paired section, we defined deleted terms as those present in the draft but absent in the final text, and added terms as those absent in the draft but present in the final text. We estimated the frequency of deletions among sections containing at least one stigmatizing term in the draft, and the frequency of introductions both overall and among sections whose drafts contained no stigmatizing terms, reporting proportions with 95% confidence intervals. We summarized deletion and addition intensity using the distributions of the numbers of removed and added terms among sections with any such change, and identified the most frequently deleted and most frequently added terms by aggregating term frequencies across sections.

## Results

From an initial 98,331 section-level draft–final pairs, we excluded 32,034 records in which the clinician finalized text was empty (missing or whitespace-only), yielding a final analytic sample of 66,297 paired sections. Using the final analytic corpus of 66,297 ambient AI drafted sections, 14,207 (21.4%) contained at least one SL mention (95% CI, 21.1% to 21.7%). In total, the mean section length was 205.4 words in the ambient AI–generated drafts and contained 19,877 raw SL mentions. The 30 most frequent stigmatizing terms observed in AI drafts are summarized in Table 1. Among sections containing at least one stigmatizing mention, the length-standardized mention density (per 200 words) had a median of 0.98 (IQR, 0.70 to 1.45). SL mentions were highly concentrated, where the top 10% of draft sections with the highest SL counts accounted for approximately 62% of all stigmatizing mentions.

**Table 1:** Top 30 SL terms most frequently appearing in ambient AI-generated drafts, sorted by decreasing frequency

| SL term | Freq. |
|---|---|
| control | 6,935 |
| believes | 3,440 |
| unsure | 1,841 |
| failure | 1,507 |
| declined | 1,222 |
| challenging | 950 |
| uncontrolled | 577 |

| SL term | Freq. |
|---|---|
| habit | 173 |
| declines | 169 |
| reluctant | 154 |
| demanding | 139 |
| alcoholic | 137 |
| morbid obesity | 127 |
| fails | 113 |

| SL term | Freq. |
|---|---|
| narcotics | 88 |
| controls | 80 |
| intolerant | 66 |
| unpleasant | 66 |
| narcotic | 62 |
| agitated | 43 |
| disruptive | 41 |

| aggressive | 377 | unpredictable | 104 | substance abuse | 41 |
|---|---|---|---|---|---|
| states | 360 | abuse | 97 | drunk | 39 |
| failed | 201 | anger | 96 | non-alcoholic | 38 |

Across the 66,297 clinician finalized sections, 15,910 (24.0%) contained at least one SL mention (95% CI, 23.7 to 24.3%). In general, the mean section length was 209.5 words, and the text contained 23,116 raw SL mentions. Among sections containing at least one mention, the length-standardized mention density (per 200 words) had a median of 0.99 (IQR, 0.68 to 1.49). The 30 most frequent stigmatizing terms observed in clinician finalized sections are summarized in Table 2. Similarly, SL mentions were also highly concentrated in clinician finalized sections, with the top 10% accounting for approximately 60% of all mentions.

**Table 2:** Top 30 SL terms most frequently appearing in clinician finalized sections, sorted by decreasing frequency

| SL term | Freq. | SL term | Freq. | SL term | Freq. |
|---|---|---|---|---|---|
| control | 6,404 | failed | 413 | fails | 113 |
| believes | 3,251 | aggressive | 356 | anger | 104 |
| states | 1,674 | reluctant | 261 | unpredictable | 93 |
| failure | 1,663 | morbid obesity | 220 | controls | 77 |
| declined | 1,639 | alcoholic | 201 | narcotics | 76 |
| unsure | 1,594 | habit | 171 | narcotic | 54 |
| apparently | 1,148 | refused | 167 | unpleasant | 52 |
| uncontrolled | 883 | intolerant | 130 | agitated | 51 |
| challenging | 833 | demanding | 116 | refuses | 45 |
| declines | 447 | abuse | 113 | substance abuse | 45 |

Across note sections, SL prevalence varied substantially, shown in Table 3. In both AI drafts and clinician finalized text, SL was most common in HPI sections (28.5% in drafts; 30.3% in final) and A&P sections (18.1% in drafts; 21.9% in final), which also comprised the largest share of the dataset. In contrast, SL was rare in Physical Exam sections (0.1% in drafts; 0.4% in final). Results sections showed intermediate prevalence but a noticeable increase from draft to final (0.6% to 2.9%).

**Table 3.** SL prevalence by note section in AI drafts and clinician finalized text

| Note section | # of total sections | # of section with ≥ 1 SL mention in draft | # of section with ≥ 1 SL mention in final |
|---|---|---|---|
| HPI | 32,553 | 9,283 (28.5%) | 9,866 (30.3%) |

| A&P | 27,138 | 4,905 (18.1%) | 5,945 (21.9%) |
|---|---|---|---|
| Physical Exam | 3,733 | 2 (0.1%) | 15 (0.4%) |
| Results | 2,873 | 17 (0.6%) | 84 (2.9%) |

To characterize how clinician editing changed SL presence from draft to final, we categorized paired sections into four states (as shown in Table 4): neither version contained SL (48,016; 72.4%), both versions contained SL (11,836; 17.9%), SL appeared only in the final (4,074; 6.1%), or only in the draft (2,371; 3.6%). McNemar's test indicated a significant asymmetry between introductions and removals ($P < 0.001$), with introductions occurring more frequently than removals. The test suggested that clinician edits were more likely to introduce than to eliminate SL at the section level. We further compared SL density between AI drafts and clinician finalized text using the Wilcoxon signed-rank test. Across all paired sections, mention density differed significantly between versions ($P < 0.001$). Since most section pairs contained no SL in either version, we focused interpretation on 18,281 sections where SL appeared in at least one of the draft or final. In this subset, the paired difference in mention density remained statistically significant ($P < 0.001$), the change in length-standardized mention density from draft to final (per 200 words) had a median of 0.03 (IQR, -0.15 to 0.53).

**Table 4:** Distribution of paired SL terms presence states

| **Paired State** | **Total number of Sections** | **Proportion of Total Sections (%)** |
|---|---|---|
| Neither | 48,016 | 72.4 |
| Both | 11,836 | 17.9 |
| Clinician final-only | 4,074 | 6.1 |
| Ambient AI draft-only | 2,371 | 3.6 |

Among 14,207 draft sections containing at least one SL mention, 3,307 (23.2%) had at least one stigmatizing term removed in the clinician finalized text (95% CI, 22.6% to 24.0%). Removal intensity was generally low, the median number of removed terms was 1 among sections with any removal. Across all paired sections, clinicians removed a total of 4,014 stigmatizing terms. Most removals occurred in HPI and A&P sections (2,234 and 1,775 removed mentions, respectively), with comparatively few removals in Physical Exam (2) and Results (3) sections. Table 5 presents the 30 most frequently removed stigmatizing terms during clinician editing.

**Table 5:** Top 30 SL terms most frequently removed during clinician editing, sorted by decreasing frequency

| **SL term** | **Freq.** |
|---|---|
| control | 1,101 |
| believes | 619 |
| unsure | 474 |
| declined | 229 |
| failure | 196 |
| challenging | 175 |

| **SL term** | **Freq.** |
|---|---|
| reluctant | 35 |
| declines | 31 |
| habit | 29 |
| alcoholic | 26 |
| fails | 26 |
| demanding | 25 |

| **SL term** | **Freq.** |
|---|---|
| narcotics | 19 |
| unpredictable | 19 |
| intolerant | 16 |
| drunk | 13 |
| refused | 12 |
| skeptical | 12 |

| aggressive | 98 |
| --- | --- |
| uncontrolled | 96 |
| states | 67 |
| failed | 38 |
| controls | 20 |
| narcotic | 20 |
| unpleasant | 20 |
| abuse | 19 |
| morbid obesity | 10 |
| non-alcoholic | 10 |
| anger | 8 |
| refusal | 8 |

Across 66,297 all paired sections, 5,453 (8.2%) introduced at least one new stigmatizing term in the clinician finalized text (95% CI, 8.0% to 8.4%). Among 52,090 sections whose drafts contained no SL, 4,074 (7.8%) introduced at least one stigmatizing term (95% CI, 7.6% to 8.1%). Likewise, introduction intensity was generally low, the median number of added terms was 1 among sections with any introduction. In total, clinician editing contributed 7,253 added raw SL mentions. Most introduced mentions occurred in HPI and A&P sections (3,656 and 3,504 introduced mentions, respectively), with comparatively few in Physical Exam (16) and Results (77) sections. Of the added raw mentions, 5,594 (77.1%) of these added mentions occurred in sections whose drafts were previously clear of SL, while 1,659 (22.9%) occurred in sections whose drafts already contained SL. Table 6 shows the 30 most frequently introduced stigmatizing terms after clinician editing.

**Table 6:** Top 30 SL terms most frequently introduced after clinician editing, sorted by decreasing frequency

| SL term | Freq. |
| --- | --- |
| states | 1,176 |
| apparently | 853 |
| control | 831 |
| declined | 593 |
| believes | 481 |
| failure | 444 |
| uncontrolled | 333 |
| declines | 294 |
| unsure | 252 |
| failed | 230 |

| SL term | Freq. |
| --- | --- |
| reluctant | 134 |
| refused | 109 |
| morbid obesity | 100 |
| alcoholic | 90 |
| aggressive | 88 |
| intolerant | 78 |
| challenging | 59 |
| abuse | 33 |
| habit | 30 |
| fails | 29 |

| SL term | Freq. |
| --- | --- |
| anger | 21 |
| refuses | 19 |
| controls | 16 |
| agitated | 14 |
| angry | 14 |
| unreliable | 13 |
| adamant | 12 |
| narcotic | 12 |
| nonadherence | 12 |
| poor historian | 11 |

**Discussion**

Across note sections, clinician editing was associated with a higher overall presence of SL in the finalized record. Overall, 21.4% of ambient AI draft sections contained at least one SL mention, compared with 24.0% of clinician finalized sections, indicating that SL was observed in a higher proportion of sections after clinician editing. The prevalence of SL within both draft and clinician finalized text are concordant with current literature [10,19,24,39,40]. This increase was most evident in the HPI and A&P sections where most clinical interpretation, judgements and labeling occur, and these patterns suggest that edits made during HPI and A&P are the key locus of added SL. Despite this shift in section-level prevalence, SL intensity among affected sections was similar across versions: the median length-standardized mention density (per 200 words) was 0.98 in drafts and 0.99 in clinician finalized text, implying that clinician edits primarily changed whether SL appeared in a new section rather than substantially increasing the within-section concentration once present. To move beyond overall prevalence, our analysis of paired draft–final sections clarifies the direction of change during clinician editing. By categorizing sections into whether SL appeared in four states, we observed a net shift toward more frequent appearance in the clinician finalized text indicating that clinician edits are more likely to introduce SL than to remove it. Complementing this presence-absence signal, the paired Wilcoxon comparisons suggest that clinician edits also alter the distribution of mention density between drafts and final text. These results formulate the same conclusion that clinician editing doesn't simply remove SL from ambient AI drafts, but can be a meaningful source of SL that enters the finalized record.

Across 66,297 paired sections, clinicians removed at least one stigmatizing term in 3,307 sections and introduced at least one new stigmatizing term in 5,453 sections. In aggregate, clinician edits removed 4,014 stigmatizing terms and introduced 7,253 new stigmatizing mentions. Among the top 30 most frequently removed and introduced stigmatizing terms, 22 overlapped, indicating that many high-frequency terms were both added and removed across different sections. The non-overlapping terms highlight asymmetries in how specific expressions were edited. Terms appearing only among the most frequently removed included "demanding," "unpleasant," "narcotics," "unpredictable," "drunk," "skeptical," "refusal," and "non-alcoholic," whereas terms appearing only among the most frequently introduced included "apparently," "refuses," "agitated," "angry," "unreliable," "adamant," "poor historian," and "nonadherence." This pattern suggests that clinician editing can both eliminate negative descriptors of personality (e.g., "drunk," "unpleasant") and more likely to introduce clinician-judged negative descriptors of behaviors of patients (e.g., "poor historian," "unreliable," "nonadherence") which are still stigmatizing but in a more subtle form, aligned with categories "Disapproval" and "Stereotyping" introduced in Park et al.'s study [11,17,19,24].

This study has several limitations. First, the data were extracted from a single health system, which may limit generalizability to institutions with different patient populations, documentation practices, or ambient AI workflows. Second, our analyses focused on differences between ambient AI drafts and clinician finalized text and did not evaluate how SL varies across patient social and demographic factors (e.g., race/ethnicity, insurance status, age, sex, or other social determinants), which will be important for understanding potential disparities. Third, we restricted the analysis to outpatient notes; SL and editing behaviors may differ in inpatient, emergency, or other documentation contexts. Fourth, we did not stratify analyses by ambient AI vendor or clinical specialty; differences in model behavior, draft-generation workflows, implementation context, and specialty-specific documentation norms may contribute to variation in SL and editing patterns. Finally, our lexicon-based approach quantified individual term mentions and did not capture contextual or sentence-level expressions of stigma, which may require complementary context-aware methods to assess.

Overall, this study provides a quantitative blueprint for measuring SL in ambient AI–generated draft notes and subsequent clinician edits, showing that clinician editing does not merely remove SL from AI drafts but can also introduce SL into the finalized record. Building on this preliminary analysis, several directions emerge for further investigating and reducing SL in ambient AI workflows. Future work should trace SL across the end-to-end pipeline, from encounter audio and transcripts to AI drafts and finalized notes, to assess where stigmatizing terms originate and where they are most likely to be removed or introduced. Analyses stratified by specialty and vendor can clarify whether patterns reflect workflow differences, documentation norms, or model behavior. We also plan to examine section-level "hot spots" and clinician-level variation to distinguish whether stigma concentration reflects a small subset of users versus a broader prevalence phenomenon, and to identify where targeted feedback or guardrails may be most effective. Finally, longitudinal comparisons across pre-AI and post-AI periods can characterize how stigma evolves with adoption and inform interventions that combine measurement with practical mitigation strategies.

**Conclusion**

In this study, we conducted a large-scale, paired comparison of SL in ambient draft notes and the corresponding clinician finalized notes using a lexicon of 251 terms from prior study [36]. At the section level, we quantified SL in paired ambient AI drafts and clinician finalized text and tested how clinician editing changed its prevalence and composition, including whether terms were removed or newly introduced. Our findings show that approximately 21% of ambient AI draft sections contained at least one SL mention, and SLwas observed in a higher proportion of sections after clinician editing. Paired statistical tests indicated that clinician edits significantly altered both the prevalence and SL density from draft to final text. Notably, introductions occurred more frequently than removals, indicating that clinician editing can be a net source of SL entering the finalized record. Further investigation is needed for drill-down analysis in end-to-end provenance, vendor and specialty differences, section and clinician hot spots, and longitudinal pre/post adoption trends.